\documentclass[conference]{IEEEtran}

\pdfoutput=1

\usepackage{amsmath,amssymb,amsfonts}
\usepackage{amsthm}
\usepackage{algorithm}
\usepackage{algpseudocode}
\usepackage{graphicx}
\usepackage{textcomp}
\usepackage{xcolor}
\usepackage{multirow}

\usepackage{amsfonts}
\usepackage{amssymb}
\usepackage{mathtools}

\usepackage{subfig}

\algnewcommand{\LeftComment}[1]{\Statex \(\triangleright\) #1}
\algblock{ParFor}{EndParFor}
\algnewcommand\algorithmicparfor{\textbf{parfor}}
\algnewcommand\algorithmicpardo{\textbf{do}}
\algnewcommand\algorithmicendparfor{\textbf{end\ parfor}}
\algrenewtext{ParFor}[1]{\algorithmicparfor\ #1\ \algorithmicpardo}
\algrenewtext{EndParFor}{\algorithmicendparfor}

\graphicspath{ {./imgs/} }

\usepackage[nolist,nohyperlinks]{acronym}
\newacro{pdf}[PDF]{probability density function}
\newacro{pmf}[PMF]{probability mass function}
\newacro{mcmc}[MCMC]{Markov Chain Monte Carlo}
\newacro{RFS}[RFS]{Random Finite Set}
\newacro{FISST}[FISST]{Finite Set Statistics}
\newacro{GLMB}[GLMB]{Generalized Labeled Multi-Bernoulli}
\newacro{LMB}[LMB]{Labeled Multi-Bernoulli}
\newacro{iid}[i.i.d]{independent and identically distributed}
\newacro{ospa2}[OSPA(2)]{optimal subpattern assignment (2)}

\newacro{JPDA}[JPDA]{Joint Probabilistic Data Association}
\newacro{MHT}[MHT]{Multiple Hypothesis Tracking}
\newacro{GNN}[GNN]{Global Nearest Neighbor}

\newacro{PHD}[PHD]{probability hypothesis density}
\newacro{CPHD}[CPHD]{cardinalized probability hypothesis density}
\newacro{CBMeMBer}[CBMeMBer]{cardinality balanced multi-target multi-Bernoulli}
\newacro{MeMBer}[MeMBer]{multi-target multi-Bernoulli}
\newacro{PMBM}[PMBM]{Poisson multi-Bernoulli mixture}

\newcommand{\x}{\textbf{x}}

\begin{document}
\title{On Gibbs Sampling Architecture for Labeled Random Finite Sets Multi-Object Tracking}

\author{
\IEEEauthorblockN{Anthony Trezza}
\IEEEauthorblockA{\textit{Electrical Eng. and Computer Sci. Dept.}\\
\textit{Syracuse University}\\
Syracuse, USA \\
attrezza@syr.edu}
\and
\IEEEauthorblockN{Donald J. Bucci Jr.}
\IEEEauthorblockA{\textit{Advanced Technology Laboratories}\\
\textit{Lockheed Martin}\\
Cherry Hill, USA \\
donald.j.bucci.jr@lmco.com}
\and
\IEEEauthorblockN{Pramod K. Varshney}
\IEEEauthorblockA{\textit{Electrical Eng. and Computer Sci. Dept.} \\
\textit{Syracuse University}\\
Syracuse, USA \\
varshney@syr.edu}
}

\maketitle
\begin{abstract}
Gibbs sampling is one of the most popular Markov chain Monte Carlo algorithms because of its simplicity, scalability, and wide applicability within many fields of statistics, science, and engineering.
In the labeled random finite sets literature, Gibbs sampling procedures have recently been applied to efficiently truncate the single-sensor and multi-sensor $\delta$-generalized labeled multi-Bernoulli posterior density as well as the multi-sensor adaptive labeled multi-Bernoulli birth distribution.
However, only a limited discussion has been provided regarding key Gibbs sampler architecture details including the Markov chain Monte Carlo sample generation technique and early termination criteria.
This paper begins with a brief background on Markov chain Monte Carlo methods and a review of the Gibbs sampler implementations proposed for labeled random finite sets filters.
Next, we propose a short chain, multi-simulation sample generation technique that is well suited for these applications and enables a parallel processing implementation.
Additionally, we present two heuristic early termination criteria that achieve similar sampling performance with substantially fewer Markov chain observations.
Finally, the benefits of the proposed Gibbs samplers are demonstrated via two Monte Carlo simulations.
\end{abstract}

\begin{IEEEkeywords}
Gibbs Sampling, Markov Chain Monte Carlo, Random Finite Sets, Multi-object Tracking
\end{IEEEkeywords}
\IEEEpeerreviewmaketitle
\section{Introduction}\label{sec::intro}


A common task for many applications in the field of statistics is to generate samples from a joint probability distribution.
The generated samples are often used to approximate the moments of the distribution, create approximate supports of the distribution for recursive estimation of the distribution, or truncate a solution space via sampling.
However in practice, most joint probability distributions are difficult to sample using direct sampling methods, challenging to compute analytically or intractable to evaluate numerically.

Several Monte Carlo sampling methods such as rejection sampling and importance sampling have been suggested to alleviate this limitation.
They work by generating samples from an easier-to-sample-from distribution known as a \textit{proposal distribution}.
These samples are then updated such that they can be considered as  \ac{iid} samples from the target joint distribution.
This update procedure typically requires an evaluation of the non-normalized target joint distribution.
If an application's target joint distribution can be evaluated up to a normalizing constant, then these procedures are very powerful tools, but in many applications evaluation of the joint distribution is untenable.
These approaches also suffer severe scaling limitations due to the curse-of-dimensionality, and require the selection of an informative and easy-to-sample-from proposal distribution, which is often difficult to do in practice \cite{Bishop2006}.

In 1953, shortly after the introduction of ordinary Monte Carlo methods, \ac{mcmc} methods were proposed by Metropolis et al. at Los Alamos National Laboratory to simulate a liquid in equilibrium with its gas phase \cite{Metropolis1953}.
Their key realization was that they did not need to simulate the exact dynamics of the system.
Instead, they could simulate a \textit{Markov chain} that converges to the desired distribution \cite{Brooks2011}.
This insight led to the discovery of the \textit{Metropolis algorithm}.
In 1970, Hastings generalized the Metropolis algorithm for non-symmetric proposal Markov transition distributions, resulting in the still-popular \ac{mcmc} procedure known as the \textit{Metropolis-Hastings algorithm} \cite{Bishop2006, Brooks2011}.
Once adopted by the statistics community in the 1990s, its applications became widespread for generating samples from challenging joint distributions since \ac{mcmc} solutions scale much better in high dimensional spaces.
However, like ordinary Monte Carlo sampling procedures, the Metropolis-Hastings algorithm still requires the evaluation of the non-normalized target joint distribution and still requires the selection of an informative and easy-to-sample-from proposal Markov transition density \cite{Brooks2011}.

Gibbs sampling is a special case of the Metropolis-Hastings algorithm where the proposal Markov transition distribution is selected as the conditional distribution of the desired target joint distribution and it iterates through sample space by sampling one random variable at a time.
In practice, the conditional distribution is often significantly easier to sample from and evaluate compared to the joint distribution since it holds all random variables constant except one.
The resulting procedure \textit{does not require evaluation of the target joint distribution}, \textit{scales well in high-dimensional spaces} and \textit{does not require the selection of a proposal distribution}.
It was originally introduced in 1984 by Geman and Geman \cite{Geman1984} with an application to image restoration, apparently without knowledge of earlier work in \ac{mcmc} methods.
Following that, Gelfrand and Smith generalized the applicability of the approach for statistical methods by expanding the work done by Geman and Geman and broadening the substitution sampling work pioneered by Tanner and Wong \cite{Tanner1987, Gelfand2000}.
As inexpensive computing resources began to flourish through 1990s, Gibbs sampling techniques subsequently exploded in popularity due to the simplicity and vast applicability of the approach.
Since then, many papers and books have been written on this topic and its applications to statistical fields such as Bayesian inference and multi-object tracking \cite{Bishop2006, Brooks2011, Casella1992, Vo2016, Trezza2022}.

The goal of multi-object tracking is to jointly estimate the number of objects and their trajectories from measurements observed at one or more sensors.
Traditional approaches to multi-object tracking independently estimate each object's random state over time and use heuristics to estimate the random number of objects in the scene.
Advances in the field of multi-object tracking reformulate the problem such that the entire scene of objects is treated as a single random variable, and estimated as a multi-object probability density known as a \ac{RFS} density.
An appropriate notion representing multi-object probability densities enables concepts such as state space modeling, Bayes recursion, and Bayes optimality to be directly translated from the single-object to the multi-object tracking case \cite{Mahler2007, Mahler2014, Vo2015}.
In \cite{Vo2013}, the concept of a labeled \ac{RFS} was proposed and the authors showed that the \ac{GLMB} distribution is a conjugate prior under the multi-object Bayes filtering recursion.
Since then, the \ac{GLMB} filter and the field of labeled \ac{RFS} filtering has grown significantly in popularity \cite{Vo2014, Vo2016, Vo2019, Reuter2014, Reuter2017}.

Within the field of labeled \ac{RFS} filtering, Gibbs sampling has been successfully applied to two key sub-problems, (1) truncating the joint prediction and update procedure for a \ac{GLMB} density \cite{Vo2016, Vo2019} and (2) truncating newborn labels from a multi-sensor adaptive birth \ac{LMB} density \cite{Trezza2022}.
These Gibbs samplers share a commonality in that their objective is to truncate a labeled \ac{RFS} distribution via \ac{mcmc} sampling by approximating a ranked assignment problem without exhaustive enumeration.
In each of these papers, a similar Gibbs sampling architecture is proposed.
In this work, we will discuss some of these Gibbs sampler architecture design details and their implications on this problem domain.
Building upon this, we will propose an alternative methodology for sample generation and two heuristic early termination criteria.

This paper is organized as follows.
Section~\ref{sec::background} provides a brief background on \ac{mcmc} methods and current Gibbs sampler architectures for labeled \ac{RFS} filters.
Section~\ref{sec::proposed} discusses the proposed sample generation approach and presents two early termination criteria.
Section~\ref{sec::simulations} provides results from two simulations depicting the benefits of the proposed Gibbs sampling approach.

\section{Background}\label{sec::background}
\subsection{Notation}
Let $\x = [x_1, \dots, x_n]^T$ be a vector of $n$ random variables characterized by the joint \ac{pdf} or \ac{pmf} $f(\x)$.
Let $\check{\x} = \{\check{\x}_1, \dots \check{\x}_\ell\}$ be a set of $\ell$ realizations of $\x$ generated in accordance with $f(\x)$.
Let $f(x_{n'} | \x_{-n'})$ denote the conditional distribution of $x_{n'}$ where $\x_{-n'}$ denotes all elements of $\x$ except $n'$.
\subsection{Markov Chain Monte Carlo Methods}\label{sec::mcmc}
The idea behind \ac{mcmc} methods is to simulate a Markov chain, $\x[1], \dots, \x[k]$, such that $p(\x[k])$ converges to the desired joint distribution $f(\x)$ that we wish to sample from.
Specifically, our objective is to design an ergodic Markov chain where the equilibrium distribution is the target joint distribution.
Due to the ergodic property of the Markov chain, as $k \rightarrow \infty$, observations of the Markov chain $\x[k] \sim f(\x)$ can be used as samples from the joint distribution.
If we can construct an intelligent walk strategy through sample space then these techniques have the potential to scale significantly better than ordinary Monte Carlo sampling methods \cite{Bishop2006, Brooks2011}.

The colloquial term \textit{burn-in} is commonly used in \ac{mcmc} sampling to denote a number of observations at the beginning of an \ac{mcmc} Markov chain simulation that can be discarded.
These observations are discarded, because until the random sequence convergences to the equilibrium distribution, observations may not be considered good samples from the target distribution.
If an application can select a good starting point for the Markov chain such that, from the beginning of (or early in) the random sequence, observations of the simulated chain are representative of samples from the equilibrium distribution, then a burn-in period is not necessary.
However, in practice, there is no theoretical analysis of Markov chain dynamics that can provide a good starting point.
Unless application specific knowledge can be leveraged, two common approaches are used to seed starting points;
(1) randomly select a starting point, or
(2) begin the next Markov chain simulation at the last observation of the previous Markov chain simulation.
Assuming the previous Markov chain converged to the equilibrium distribution, then approach (2) can significantly reduce the required burn-in period since the starting value can already be considered a sample \cite{Brooks2011}.
However, it fails in unreachable Markov chains and can lead to the simulated Markov chain getting stuck in areas of high-probability due to its lack of emphasis on explorability of the sample space.
In applications where this is of concern, approach (1) can provide a more reliable starting point selection technique at the cost of a longer burn-in period.

The rate at which an ergodic Markov chain converges to the equilibrium distribution is known as the \textit{convergence rate}.
If a Markov chain has a fast convergence rate, then observations of the simulated Markov chain will quickly converge to the equilibrium distribution and hence only a few burn-in observations will be evaluated \cite{Bishop2006, Brooks2011}.

\begin{figure}[t!]
    \centering
    \includegraphics[width=0.4\textwidth]{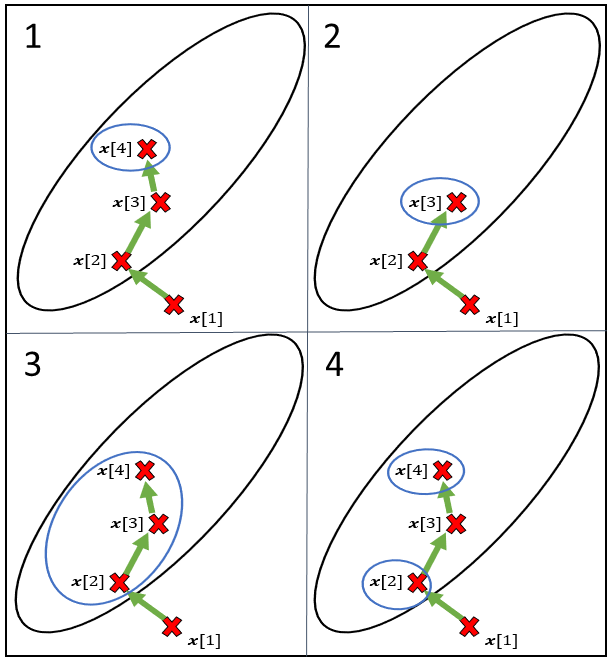}
    \caption{Four approaches for generating samples from a Markov chain simulation in accordance with target joint distribution. Selected Markov chain observations to be used as samples are highlighted by a blue ellipse.}\label{fig::mcmc_sample}
\end{figure}

Several approaches have been proposed for generating samples from the target joint distribution using the observations of the Markov chain.
Figure~\ref{fig::mcmc_sample} depicts four common techniques.
Approach 1 (top left), is the simplest approach that works by simulating the Markov chain long enough such that the final observation of the Markov chain has converged to the equilibrium distribution.
The final Markov chain observation is then used as a sample from the equilibrium distribution and the rest are discarded.
A new simulation occurs for each sample that the user wishes to generate.
Note that selecting a chain length that is ``long enough" is often difficult to do in practice and depends upon the convergence rate.
Approach 2 (top right), is similar to approach 1 in that only the final observation of the chain is used as a sample from the equilibrium distribution.
However, it employees an early termination criteria by  monitoring the convergence of the sequence.
If the sequence appears to have converged, then the chain simulation can be terminated early.
Several convergence monitoring methods have been proposed, but they can be computationally complex and none are foolproof \cite{Casella1992}.
Approach 3 (bottom left), works by simulating the chain ``long enough" to ensure convergence and then returning all of the samples as \textit{non-\ac{iid}} samples from the equilibrium distribution after discarding burn-in (if applicable).
These samples are not \ac{iid} since there is clearly a dependency upon the value of the previous sample due to the Markov property.
In applications where \ac{iid} samples are not required, this approach can be very computationally efficient since a single chain simulation can return multiple samples from the equilibrium distribution.
However, this approach runs the risk of getting stuck in high-probability regions for very long periods of time.
Finally approach 4 (bottom right), works similar to approach 3 but every $r$th observation is used as a sample from the equilibrium distribution.
For a large-enough value of $r$, these samples can be reasonably approximated as \ac{iid}.
However, similar to approach 3, although more efficient, it again runs the risk of getting stuck in regions of high probability \cite{Casella1992}.

The idea of Gibbs sampling is to update the Markov chain by replacing \textit{a subset of the random variables at a time} (usually 1) for the next observation's random vector by generating a sample from the conditional distribution, conditioned on the most recent values of all remaining variables.
That is, the Markov transition density in a Gibbs update is selected as the conditional distribution, $f(x_{n'} | \x_{-n'})$.
This results in taking gradual steps along the Markov chain such as, $x_1[k'-1]$ to $x_1[k']$, then $x_2[k'-1]$ to $x_2[k']$ and so forth, rather than direct steps from $\x[k'-1]$ to $\x[k']$.
The order in which the random variables are sequenced can be sequential or shuffled on each iteration to encourage exploration.
In practice, the conditional distribution is often significantly easier to sample from compared to the joint distribution since it holds all random variables constant except one  \cite{Bishop2006, Brooks2011, Geman1984, Gelfand2000, Casella1992}.


\subsection{Gibbs Sampling in Labeled \ac{RFS} Filters}\label{sec::rfs_app}
Gibbs sampler implementations for the single-sensor joint prediction and update, multi-sensor joint prediction and update, and multi-sensor adaptive birth procedures are provided in \cite[Algorithm 1]{Vo2016}, \cite[Algorithm 1, 2]{Vo2019} and \cite[Algorithm 1]{Trezza2022}, respectively.
These Gibbs samplers share a commonality in that their objective is to truncate a labeled \ac{RFS} distribution via \ac{mcmc} sampling to approximate the ranked assignment problem without exhaustive enumeration.
Due to this, all simulated Markov chain observations can be accepted as samples since all solutions can contribute to reducing truncation error, thus removing the need for sample burn-in rejection.
Since each of these samplers exhibit an exponential convergence rate, the burn-in period of a Markov chain simulation is expected to be short.

Although ergodic and can be initialized at any point, these samplers do have an intuitive initialization point from which the sampler can easily maneuver to any region of sample space.
In the single and multi-sensor joint prediction and update Gibbs samplers, this initialization point is the all missed detection hypothesis and for multi-sensor adaptive birth, this initialization point is the all missed detection multi-sensor measurement tuple.
In both cases, this initialization point has the least explicit or implicit constraints on its next Markov chain observation and is connected to all possible solutions.
In fact, this point is used to prove the irreducibility of each Gibbs sampler \cite{Vo2016, Vo2019}.
Alternatively, as described in \cite{Vo2016}, the highest weighted solution can be used as an initialization point by solving an optimal assignment problem.
In practice, this procedure is often not necessary as the highest weighted solution has a high likelihood of being observed once the sampler converges to the target distribution.

The sample generation approach proposed in these Gibbs samplers is a single-simulation approach using one long Markov chain (i.e., Figure~\ref{fig::mcmc_sample}, Approach 3).
That is, one Markov chain is initialized and updated, accepting every observation as a sample, until the desired number of samples have been generated.
Since these Markov chains are irreducible, simulating a single Markov chain is a valid approach since every state can eventually be reached from any other state.
However, in practice, although it may be \textit{possible} to reach any other state, the likelihood of actually doing so can be incredibly small and require a large number of samples.
This occurs when a Markov chain simulation gets stuck in a high probability island with a very low saddle.
In the context of the single or multi-sensor joint prediction and update, this can be seen when you have high detection probability, and as a result would need the chain go through multiple, very low probability transitions back to the all-missed-detection state to go in-between association decisions.

Annealing and tempering procedures are briefly mentioned in \cite{Vo2016, Vo2019, Trezza2022} as techniques to encourage more diverse regions of sample space.
Such approaches are especially important for long Markov chains that risk getting caught in high probability islands due to pseudo-convergence.
Additionally, \cite{Trezza2022} suggests randomizing the order that the Gibbs update procedure iterates through the random variables.
This is a common procedure \cite{Bishop2006} and can significantly reduce the probability of a Markov chain simulation getting stuck in a high-probability cycle.

\section{Proposed Gibbs Sampling Approach for Labeled \ac{RFS} filters}\label{sec::proposed}
We propose to use a multi-simulation sample generation approach using short Markov chains as shown in Algorithm~\ref{alg::gibbs}.
That is, rather than simulating a single, long Markov chain and using its observations as samples, we propose simulating many short Markov chains in parallel and using the concatenation of their observations as samples.
This provides several favorable properties over the single Markov chain simulation approach.
First, the Gibbs samplers under consideration have an exponential convergence rate meaning the chain length does not need to be long for a simulation to converge to the target distribution.
Second, burn-in observations are not discarded in a truncation procedure, meaning the initial observations at the beginning of the Markov chain are not wasteful.
Third, it reduces the chances that any single simulation will get stuck in a high probability island for a long period of time, allowing for a higher chance to explore diverse regions of the sample space.
Finally, since each Markov chain is simulated independently, it enables them to be executed in parallel.

A multi-short run sample generation architecture is not a novel concept and has been proposed in many Gibbs sampling applications.
The authors in \cite{Brooks2011} heed warning against using a multi-short run simulation.
Their rationale is that it can be alluring to gravitate towards short Markov chains to alleviate the phenomenon of pseudo-convergence.
This is discussed in the context of using \ac{mcmc} sampling in a black-box scenario (i.e., you are given a Markov chain and you know nothing about the transition probabilities of the Markov chain, the invariant distribution, the chain dynamics, etc.).
In this regard, we agree that if nothing is known about the \ac{mcmc} simulation or if the invariant distribution is difficult to reason about, then it is challenging to know what an appropriate chain length should be.
However, the labeled \ac{RFS} Gibbs samplers discussed here are used for truncation by approximating a ranked assignment solution.
In this context, the Markov chain is well understood and intuitive.
Sampled solutions can be directly compared against the finitely many solutions to evaluate sampling performance.
Using this, one can easily gain confidence into appropriate chain lengths required to achieve a desired performance on average.

\begin{algorithm}[t!]
\caption{Proposed Short Chain, Multi-simulation Gibbs Sampler and Early Termination Criteria}\label{alg::gibbs}
	\begin{algorithmic}[1]
		\renewcommand{\algorithmicrequire}{\textbf{Input:}}
		\renewcommand{\algorithmicensure}{\textbf{Output:}}
		\Require
			\Statex $\x[0]$, $N_S$, $N_C$, $\tau_{\textit{stall}}$, $\tau_{\textit{stale}}$
		\Ensure
			\Statex $\{\check{\x}_1, \dots, \check{\x}_{\ell}\}$
        %
		\For{$i = 1, \dots, N_S$}
            %
            %
    	    \State $\check{\x}^{(i)}[0] = \x[0]$
            \For{$j = 1, \dots, N_C$}
                \State $\check{\x}^{(i)}[j] = \text{GibbsUpdate}(\check{\x}^{(i)}[j - 1])$
                %
                \State $n_u = \text{NumUnique}(\check{\x}^{(i)}[0:j])$
                \If{$j - n_u \ge \tau_{\textit{stall}}$}
                    \State break
                \EndIf
            \EndFor
            %
			\State $\check{\x} = \check{\x} \cup \check{\x}^{(i)}$
            \State $n_u = \text{NumUnique}(\check{\x})$
            \If{$|\check{\x}| - n_u \ge \tau_{\textit{stale}}$}
                \State break
            \EndIf
		\EndFor
	\end{algorithmic}
\end{algorithm}

An additional benefit of a multi-short run simulation is that it lends itself to more natural early termination criteria.
In this work, we propose two heuristic early termination criteria for the multi-short run sample generation technique; the \textit{stall} and \textit{stale} criteria.
The stall criterion is a simple check for pseudo-convergence of a single Markov chain.
That is, if a Markov chain simulation observes $\tau_{\textit{stall}}$ number of non-unique solutions within a single chain, then that Markov chain simulation is terminated.
The stale criterion is another simple check to see if multiple completed Markov chain simulations have produced no unique solutions.
That is, if $\tau_{\textit{stale}}$ number of completed Markov chain simulations have not contributed any unique solutions, then terminate the entire \ac{mcmc} sampler.
Both heuristics are tightly aligned with the ranked assignment problem domain and do not generalize well for arbitrary Gibbs samplers.
As we will demonstrate in Section~\ref{sec::simulations}, these two early termination criteria perform exceedingly well for various multi-object tracking scenarios.

Note that if the stale early termination criterion is not considered (lines 11-14), Algorithm~\ref{alg::gibbs} can be trivially parallelized by executing each simulation in a separate thread.
The benefit of parallelization greatly depends upon the amount of time that each Gibbs update takes and the amount of overhead required for thread management.

\section{Simulations}\label{sec::simulations}
In this section we focus the performance evaluation of the proposed Gibbs sampler, without consideration of the parallel processing implementation, applied to the single-sensor ranked assignment problem \cite{Vo2016}.
Similar results are expected for the multi-sensor ranked assignment and multi-sensor adaptive birth Gibbs samplers.
Results of two simulation experiments are presented to demonstrate the performance improvements of the proposed approaches;
the first (Section \ref{sec::sim::eff}) analyzes the sampling performance using manually generated ranked assignment cost matrices,
and the second (Section \ref{sec::sim::tracking}) investigates the multi-target tracking performance impact when incorporated into a labeled \ac{RFS} filter using a common tracking example dataset.
In all simulations, the results are compared against the original Gibbs sampler implementation.

\subsection{Sampling Performance Simulation}\label{sec::sim::eff}
%
\begin{figure}[t!]%
    \centering
    \subfloat[Uniform Cost Matrices]{\includegraphics[width=0.4\textwidth]{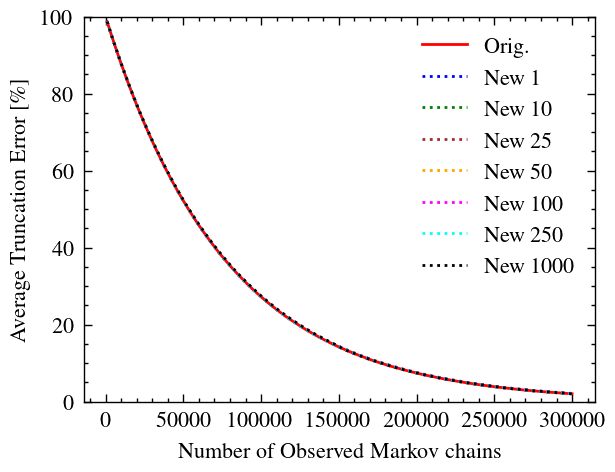} }%
    \qquad
    \subfloat[Random Cost Matrices]{\includegraphics[width=0.4\textwidth]{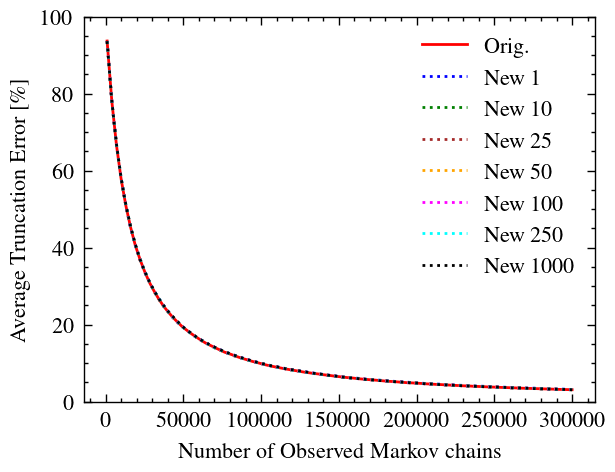} }%
    \qquad
    \caption{Gibbs sampling scalability and truncation performance as the number of Markov chain observations increases. ``Orig." denotes the original Gibbs sampling implementation and ``New X" denotes the proposed implementation where ``X" denotes the Markov chain length. Results are averaged over $100$ Monte Carlo iterations.}%
    \label{fig::sim::sweep}%
\end{figure}

In this simulation, the $\delta$-\ac{GLMB} ranked assignment cost matrices are manually constructed in the form provided in Figure 1 of \cite{Vo2016}.
Briefly, the cost matrix $C$ is given by, 
\begin{equation}\label{eq::costmatrix}
    C_{ij} = \begin{cases}
        -\ln(\eta_i(j)) & j\in\{1:M\}\\
        -\ln(\eta_i(0)) & j = M + i \\
        -\ln(\eta_i(-1)) & j = M + P + i \\
        \infty & \text{otherwise}
    \end{cases}
\end{equation}
where,
\begin{equation}
    \eta_i(j) = \begin{cases}
        1 - \bar{P}_s^{(\xi)}(\ell_i) & 1 \leq R, j < 0\\
        \bar{P}_s^{(\xi)}(\ell_i)\bar{\psi}^{(\xi, j)}_{Z_+}(\ell_i) & 1 \leq R, j \geq 0\\
        1-r_{B,+}(\ell_i) & R+1 \leq i \leq P, j < 0 \\
        r_{B,+}(\ell_i)\bar{\psi}^{(\xi, j)}_{Z_+}(\ell_i) & R + 1 \leq i \leq P, j \geq 0.
    \end{cases}
\end{equation}
The number of measurements, $M$, number of components from the last time-step, $R$, and the number of newborn components, $B$, determines the size of cost matrix $P\times(M+2P)$ where $P = R+B$.
Rows $1$ through $R$ and $R+1$ through $P$ in Equation~(\ref{eq::costmatrix}) represent the persisting and the newborn label segments of the cost matrix respectively.
Columns $1$ through $M$, $M+1$ through $M+P$, and $M+P+1$ through $M+2P$ represent the the survived and detected, survived and missed detection, and died or not born segments of the cost matrix respectively.
For label $\ell_i$ in $\delta$-\ac{GLMB} component $(\xi, I)$, 
$\bar{P}_s^{(\xi)}(\ell_i) = r(\ell_i)P_s$ is the average survival probability,
$P_s$ is the constant survival probability, 
$r(\ell_i)$ is the existence probability of label $\ell_i$, $r_{B,+}$ is the birth probability, 
and $\bar{\psi}^{(\xi, j)}_{Z_+}(\ell_i)$ is the average pseudolikelihood given measurement $Z_+$ given by,
\begin{equation}
    \bar{\psi}^{(\xi, j)}_{Z_+}(\ell_i) = \begin{cases}
        \bar{P}_d^{(\xi)}(\ell_i)q(z_j|x, \ell_i) & j \in \{1, \dots, |Z_+|\}\\
        1-\bar{P}_d^{(\xi)}(\ell_i) & j = 0
    \end{cases},
\end{equation}
where $\bar{P}_d^{(\xi)}(\ell_i) = r(\ell_i)P_d$ is the average probability of detection,
$P_d$ is the constant probability of detection, 
$q(z_j|x, \ell_i) = g(z_j|x,\ell_i)/\kappa(z_j)$ is the single-object measurement likelihood function normalized by the clutter intensity.
The cost of an assignment matrix $S$ is given by,
\begin{equation}
    \text{tr}(S^TC) = \sum\limits^P_{i=1}\sum\limits^{M+2P}_{j=1} C_{ij}S_{ij}.
\end{equation}
For more information, see \cite{Vo2016}.

Three types of ranked assignment cost matrices are created and evaluated against, denoted the \textit{diagonal}, \textit{uniform} and \textit{random} cost matrices.
Each cost matrix will consider $R=4$ existing components, no newborn components ($B=0$), and $M=16$ measurements (i.e., cost matrices of size $4\times24$).

The diagonal cost matrix set is a trivial ranked assignment cost matrix such that each target has a measurement that uniquely corresponds to it with very high likelihood.
This is representative of a scenario with high accuracy and high detection probability sensors tracking targets that are greatly separated in the measurement space.
The cost matrix was generated deterministically such that every existing component was parameterized by 
$r(\ell_i) = 0.99$,
$P_s = 0.99$,
$P_d = 0.99$, 
$q(z_j|x, \ell_i) = 50$ when $z_j$ was generated by a real target, 
and $q(z_j|x, \ell_i) = 1e-3$ when $z_j$ was generated by clutter.

The uniform cost matrix set is a more difficult ranked assignment problem which is meant to be representative of high uncertainty scenarios such as those with low accuracy and low detection probability sensors tracking closely spaced targets in measurement space.
These cost matrices are generated deterministically by setting every feasible entry in the ranked assignment cost matrix equal to 0.5.

The random cost matrix is randomly generated at each Monte Carlo iteration and is meant to test the robustness of the Gibbs samplers against a variety of possible tracking scenarios.
It uniformally samples a $r(\ell_i)$, $P_s$ and $P_d$ between $[1e-3, 1]$, and $q(z_j|x, \ell_i)$ between $[0, 50]$.

The performance is quantified via the ranked assignment truncation error and number of Markov chain observations.
To compute the truncation error, we first generate all $M$ feasible ranked assignment solutions and compute their costs $c_f = \{c^{(1)}_f, \dots, c^{(M)}_f\}$.
Then, the costs of the $n$ unique sampled solutions are computed $c_s = \{c^{(1)}_s \dots, c^{(n)}_s\}$.
The truncation error is defined as one minus the fraction of the total cost that was sampled by the Gibbs sampler,
\begin{equation}\label{eq::eff}
    \epsilon (c_s) = 1 - \frac{\sum\limits_{i=1}^{n} c^{(i)}_s}{\sum\limits_{j=1}^{M} c^{(j)}_f}.
\end{equation}

Figure~\ref{fig::sim::sweep} shows the Gibbs sampling truncation error for each ranked assignment cost matrix as a function of the number of Markov chain observations, averaged over $100$ Monte Carlo iterations.
The proposed approach was evaluated with no early termination criteria and the Markov chain length was varied.
To maintain a consistent number of Markov chain observations, the number of simulated Markov chains was computed as the desired number of Markov chain observations divided by the Markov chain length of each simulation.
As seen in the figure, the scalability of the proposed approach is nearly identical compared to the original method.
Additionally, due to the exponential convergence rate of the ranked assignment Gibbs sampler, the Gibbs truncation performance is unaffected ($\pm 0.1\%$) by the Markov chain length.
For brevity, the results from the diagonal cost matrices are not depicted since all Gibbs samplers achieved nearly $0\%$ error with a small number of Markov chain observations.


Table~\ref{tab::sim::results} shows the percent change in the number of Markov chain observations ($\Delta \text{Obs.}$) and the absolute difference in truncation error ($\Delta \text{Error}$) compared to the original Gibbs sampler on all three ranked assignment cost matrices.
The percent change in the number of Markov chain observations was computed as $\Delta \text{Obs.} = 100 \cdot (N_o - N_s)/N_o$ where $N_o$ and $N_s$ are the number of Markov chain observations when using the original and proposed method under test respectively.
The absolute difference in truncation error was computed as $\Delta \text{Error} = |\epsilon(c_o) - \epsilon(c_s)|$, where $\epsilon(c_o)$ and $\epsilon(c_s)$ are the truncation errors when using the original and proposed method under test respectively.
The proposed techniques are abbreviated as, 
\textit{Stall}, \textit{Stale}, or \textit{Full} when only the stall, stale, or both early termination criteria was used.
Using the proposed Gibbs sampling architecture with no early termination criteria resulted in exactly the same values as the original Gibbs sampler architecture and thus was omitted from the table for clarity.
The original Gibbs sampler was parameterized using a single Markov chain simulation with a chain length of $250,000$ observations.
As seen in Figure~\ref{fig::sim::sweep}, a maximum number of $250,000$ observations was chosen since, on average, it should be sufficient to obtain at least $~95\%$ sampling efficiency for all test sets using the original Gibbs sampler.
The proposed Gibbs sampler was parameterized using $10,000$ Markov chain simulations with a Markov chain length of $25$ observations per simulation (note that this results in the same number of maximum observations as parameterized in the original Gibbs sampler).
When applicable, the early termination criteria thresholds were set as $\tau_{\textit{stall}} = 5$ observations and $\tau_{\textit{stale}} = 25$ simulations.
Results are averaged over $100$ Monte Carlo observations.

\begin{table}[]
\centering
\caption{
    Percent change in number of Markov chain observations and the absolute difference in truncation error compared to the original Gibbs sampler.
    A higher value for $\Delta$ Obs. (i.e., fewer observations than the original) and a lower value for $\Delta$ Error (i.e., similar error as the original) is better.
}
\label{tab::sim::results}
\begin{tabular}{llcc}
                       &
                       & \multicolumn{1}{l}{\textbf{$\Delta$ Obs.}} & \multicolumn{1}{l}{\textbf{$\Delta$ Error}} \\ \cline{3-4} 
\parbox[t]{2mm}{\multirow{3}{*}{\rotatebox[origin=c]{90}{Diag.}}} 
                       & \multicolumn{1}{l|}{\textbf{Stall}} & \multicolumn{1}{c|}{\textit{79.95\%}}   & \multicolumn{1}{c|}{\textit{0.04\%}}            \\ \cline{3-4} 
                       & \multicolumn{1}{l|}{\textbf{Stale}} & \multicolumn{1}{c|}{\textit{99.29\%}}    & \multicolumn{1}{c|}{\textit{0.27\%}}            \\ \cline{3-4} 
                       & \multicolumn{1}{l|}{\textbf{Full}}  & \multicolumn{1}{c|}{\textit{99.93\%}}      & \multicolumn{1}{c|}{\textit{0.40\%}}            \\ \cline{3-4} 
\\\hline \\
\cline{3-4}
\parbox[t]{2mm}{\multirow{3}{*}{\rotatebox[origin=c]{90}{Unif.}}} & \multicolumn{1}{l|}{\textbf{Stall}} 
                       & \multicolumn{1}{c|}{\textit{40.27\%}}  & \multicolumn{1}{c|}{\textit{10.46\%}}           \\ \cline{3-4} 
                       & \multicolumn{1}{l|}{\textbf{Stale}} & \multicolumn{1}{c|}{\textit{0.00\%}}  & \multicolumn{1}{c|}{\textit{0.00\%}}            \\ \cline{3-4} 
                       & \multicolumn{1}{l|}{\textbf{Full}}  & \multicolumn{1}{c|}{\textit{40.23\%}}  & \multicolumn{1}{c|}{\textit{10.45\%}}           \\ \cline{3-4} 
\\\hline \\
\cline{3-4}
\parbox[t]{2mm}{\multirow{3}{*}{\rotatebox[origin=c]{90}{Rand.}}} 
                       & \multicolumn{1}{l|}{\textbf{Stall}} & \multicolumn{1}{c|}{\textit{56.97\%}}  & \multicolumn{1}{c|}{\textit{6.34\%}}            \\ \cline{3-4} 
                       & \multicolumn{1}{l|}{\textbf{Stale}} & \multicolumn{1}{c|}{\textit{0.00\%}}  & \multicolumn{1}{c|}{\textit{0.00\%}}            \\ \cline{3-4} 
                       & \multicolumn{1}{l|}{\textbf{Full}}  & \multicolumn{1}{c|}{\textit{56.96\%}}  & \multicolumn{1}{c|}{\textit{6.35\%}}            \\ \cline{3-4} 
\end{tabular}
\end{table}

On the diagonal cost matrix test set (Table~\ref{tab::sim::results}, top row), the proposed Gibbs sampler achieved a very similar truncation error but used $99.93\%$ fewer Markov chain observations.
In this example, the major contributor to this improvement was the stale early termination criteria.
This is due to the fact that the diagonal cost matrices only contain a small number of high weight solutions, resulting in many Markov chain simulations producing no unique samples.

On the uniform cost matrix test set (Table~\ref{tab::sim::results}, middle row), the proposed Gibbs sampler increased the truncation error by $10.45\%$ but used $40.23\%$ fewer Markov chain observations.
The increase in truncation error is due to the uniform cost of every unsampled solution.
Since every valid solution in the uniform cost matrix has identical weight, every unsampled solution contributes substantially to the truncation error.
In this test set, the stall threshold is what leads to the vast majority of early terminations and thus the increased truncation error.
Since the association problem was extremely ambiguous, it is unlikely that an entire Markov chain simulation would execute without finding at least one new sample.
However, while progressing through a Markov chain simulation it is likely that some simulations would continue to observe non-unique values in its Markov chain since every transition has equal weight.

On the random cost matrix test set (Table~\ref{tab::sim::results}, bottom row), the proposed Gibbs sampler increased the truncation error by $6.35\%$ but used $56.96\%$ fewer Markov chain observations.
Similar to the uniform cost matrix, the reduced number of iterations is mainly attributed to the stall early termination criteria.

These results imply that more simulations of shorter Markov chains are favorable over fewer simulations of long Markov chains when computing each observation is computationally expensive.
The stale early termination criterion has the most utility in low-ambiguity ranked assignment problems where many Markov chain simulations are expected to generate no new samples.
This may occur in applications with high accuracy sensors with greatly spaced targets.
The stall early termination criterion can provide utility in a wider variety of scenarios.
In ranked assignment problems with many local maximum, the stall  criterion can help prevent many unnecessary observations of the same few solutions.
Similarly, the stall criterion can be useful in ranked assignment problems with high uncertainty.

\subsection{Tracking Scenario Simulation}\label{sec::sim::tracking}
\begin{figure}[t!]
    \centering
    \includegraphics[width=0.4\textwidth]{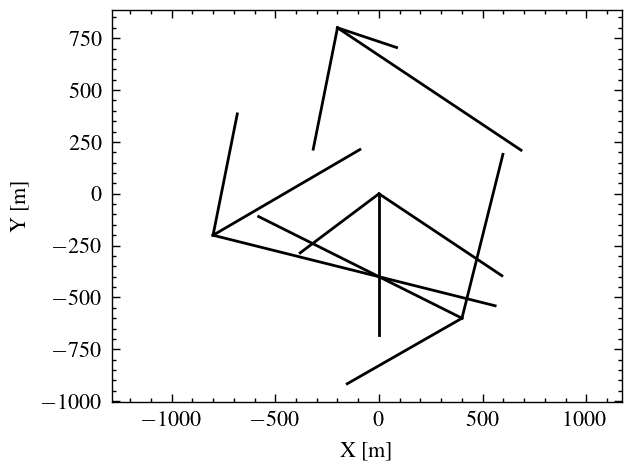}%
    \caption{Target trajectories.}%
    \label{fig::sim::vo2014_traj}
\end{figure}

\begin{figure}[t!]%
    \centering
    \subfloat[\ac{ospa2} results ]{\includegraphics[width=0.4\textwidth]{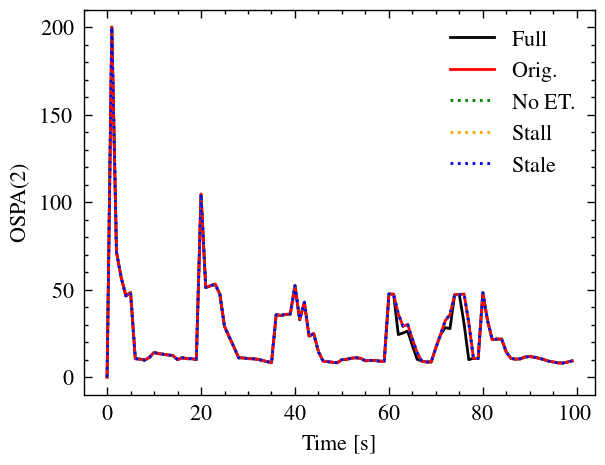} }%
    \qquad
    \subfloat[Cumulative sum of Markov chain observations]{\includegraphics[width=0.4\textwidth]{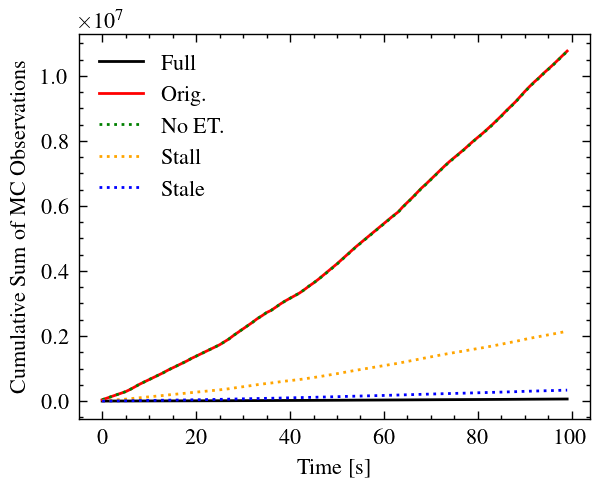} }%
    \qquad
    \caption{Multi-object tracking performance using the original and proposed ranked assignment Gibbs samplers.}%
    \label{fig::sim::vo2014}%
\end{figure}

%
In this simulation, the Gibbs sampling performance is compared using a similar multi-target tracking scenario as provided in Example 1 of \cite{Vo2007}.
The target trajectories are shown in Figure~\ref{fig::sim::vo2014_traj}.
A \ac{LMB} filter was used to propagate and update the multi-object state estimate as described in \cite{Reuter2014}.
The state space comprised of planar 2D position and velocity, $x = [p_x, \dot{p}_x, p_y, \dot{p}_y]^T$.
The targets followed a constant velocity transition model, $f_+(x_+|x) = \mathcal{N}(x_+; Fx, Gw)$ where the transition matrix and process noise matrix were given by,
\begin{equation*}
    F =
    \begin{bmatrix}
            1 & \Delta_t\\
            0 &     1
    \end{bmatrix},\qquad
    G = 
    \begin{bmatrix}
        \frac{\Delta_t^2}{2}\\
        \Delta_t
    \end{bmatrix},
\end{equation*}
respectively and $\Delta_t$ is the discrete-time sampling interval \cite{Li2003}.
The simulation was run for 100 seconds with $\Delta_t = 1$ second.
The discrete-time x and y acceleration white noises were, $w = [5, 5]^T\;m/s^2$.
The survival probability for each target was set to $p_s(x,l) = 0.99$.

A static birth procedure was used to generate a newborn \ac{LMB} distribution containing 4 labeled birth components at each timestep.
The birth probability of each component was $r_{B,+}(l_+) = 0.03$.
The spatial distribution of each birth component was modeled as a Gaussian distribution such that component label $l_+^{(i)}$'s distribution is modeled as $p_{B,+}(x_+, l^{(i)}_+) = \mathcal{N}(x_+; m^{(i)}_{B,+}, P^{(i)}_{B,+})$, with
\begin{align*}
    m^{(1)}_{B,+} &= \{0, 0, 0, 0\}, & m^{(2)}_{B,+} &= \{400, 0, -600, 0\},\\
    m^{(3)}_{B,+} &= \{-800, 0, -200, 0\}, & m^{(4)}_{B,+} &= \{-200, 0, 800, 0\},
\end{align*}
and $P^{(i)}_{B,+} = \text{diag}(\sigma_{p}^2, \sigma_{\dot{p}}^2, \sigma_{p}^2, \sigma_{\dot{p}}^2)$ with $\sigma_{p} = 300~m$ and $\sigma_{\dot{p}} = 10~m/s$ for all components.

A single linear XY-position sensor observed the targets with a constant detection probability of $p_D^{(s)}(x,l) = 0.98$.
If an object was detected, the XY-position measurement $z^{(s)} = [p_x, p_y]^T$ was observed according to the single-target measurement likelihood $g(z^{(s)}|x) = \mathcal{N}(z^{(s)}; H^{(s)}x, R^{(s)})$ with, $R^{(s)} = \text{diag}(10^2, 10^2)$ and,
$H^{(s)} = \begin{bsmallmatrix}1 & 0\\0 & 1\end{bsmallmatrix} \otimes \begin{bsmallmatrix}1 & 0\\0 & 0\end{bsmallmatrix}$ where $\otimes$ denotes the Kronecker product.
Clutter was modeled as Poisson distributed with intensity $\kappa^{(s)}(\mathbb{Z}^{(s)}) = \lambda_c^{(s)} \mathcal{U}(\mathbb{Z}^{(s)})$
where $\mathcal{U}(\mathbb{Z}^{(s)})$ is the uniform distribution over $\mathbb{Z}^{(s)}$, and
$\lambda^{(s)}_c = 50$.

This sensor model with a high clutter rate lends itself to a challenging data association problem for evaluating the ranked assignment Gibbs samplers.
The cardinality and state estimation accuracy performance was quantified using the \ac{ospa2} metric \cite{Beard2020}.
The \ac{ospa2} metric was computed using a distance cutoff value of $200~m$, a distance order of $1.0$, a sliding window length of $5$, and an expanding window weight power of $0$.

State extraction was conducted using a hysteresis existence probability-based approach as described in \cite{Reuter2014} with an upper and lower existence probability bounds set as, $\vartheta_{u} = 0.9$ and $\vartheta_{l} = 1e-3$ respectively.
The original Gibbs sampler was parameterized using a single Markov chain simulation with a chain length of $10,000$ observations.
The proposed Gibbs sampler was parameterized using $400$ Markov chain simulations with a Markov chain length of $25$ observations per simulation (note that this results in the same number of maximum observations as parameterized in the original Gibbs sampler).
When applicable, the early termination criteria thresholds were set as $\tau_{\textit{stall}} = 5$ observations and $\tau_{\textit{stale}} = 10$ simulations.

Figure~\ref{fig::sim::vo2014} (a) shows the \ac{ospa2} results.
As shown, the proposed Gibbs sampler achieved an almost identical \ac{ospa2} results as the original sampling approach.
Figure~\ref{fig::sim::vo2014} (b) shows the cumulative sum of the Markov chain observations for the duration of the simulation.
As shown, the proposed Gibbs sampler made significantly fewer Markov chain observations than the original Gibbs sampler, resulting in a $99.4\%$ reduction in the number of observations.
The original Gibbs sampler made $10,760,000$ observations as opposed to the proposed Gibbs sampler which only made $63,474$ samples over the duration of the simulation.

\section{Conclusion}\label{sec::conclusions}
In this paper, we reviewed the application of Gibbs sampling in labeled \ac{RFS} tracking and proposed an alternative sample generation method that enables parallel processing.
Additionally, we presented two early termination criteria for the proposed Gibbs sample generation approach.
We then showed via two Monte Carlo simulations that the proposed Gibbs sample generation technique and early termination criteria result in similar sampling performance as the original approach using substantially fewer Markov chain observations.
Future work is being conducted to investigate the effects of Gibbs tempering and herding on the various Gibbs samplers in the \ac{RFS} literature.

\bibliographystyle{IEEEtran}  
\bibliography{IEEEabrv, sections/ms.bib}

\end{document}